\documentclass[dvips]{acta}
\usepackage{supertabular,lscape,epsfig}
\usepackage{amssymb}
\usepackage{amsmath}
\SetPages{1}{9}
\SetVol{66}{2016}

\def \PASJ {{\it Publ.Astr.Soc.Japan}\/} 

\begin{document}

\begin{Titlepage}

\Title { Superhumps and their Evolution during Superoutbursts }

\Author {J.~~S m a k}
{N. Copernicus Astronomical Center, Polish Academy of Sciences,\\
Bartycka 18, 00-716 Warsaw, Poland\\
e-mail: jis@camk.edu.pl }

\Received{  }

\end{Titlepage}

\Abstract { Light curves of superhumps and their evolution during superoutbursts 
are analyzed by decomposing them into their Fourier components, 
including the fundamental mode and the first three overtones. 
The amplitudes of the fundamental mode are found to decrease significantly during 
superoutburst while those of the overtones remain practically constant.  
The phases of maxima of the fundamental mode {\it increase} systematically 
during superoutburst while those of the overtones -- systematically {\it decrease}. 
The combination of the two effects is responsible for the characteristic 
evolution of superhump light curves: the appearance and growth of the 
secondary humps and the spurious phase jumps in the (O-C) diagrams.  

Two intrepretations are possible. Either that instead of just one superhump 
period $P_{sh}$ there are four periods $P_k$ which resemble -- but are 
significantly different from -- the fundamental mode and the first three overtones 
of $P_{sh}$. Or -- more likely -- that those time-dependent phase shifts are genuine. 
} 
{binaries: cataclysmic variables, stars: dwarf novae }

%Sec.1
\section { Introduction } 

Superhumps are periodic light variations, with periods which are slightly longer 
than the orbital periods, observed (1) almost always in dwarf novae during their 
superoutbursts (cf. Kato et al. 2009, 2010, 2012), (2) occassionally in dwarf novae 
at quiescence and during normal outbursts (e.g. Patterson et al. 1995, 
Still et al. 2010), and (3) in the so-called permanent superhumpers -- 
the nova-like cataclysmic binaries with stationary accretion (cf. Patterson 1999). 

During the last 40 years since their discovery (Vogt 1974, Warner 1975) 
the superhumps -- particularly those observed during superoutbursts -- 
were subject to many investigations. Most of them have concentrated primarily 
or even exclusively on the superhump periods. 
Fortunately there are also papers presenting full description of superhumps, 
including their light curves and showing the evolution of their shapes during 
superoutbursts. 
Generally the superhump amplitudes decrease during superoutburst 
(cf. Smak 2010 and references therein) and their shapes vary considerably  
(e.g. Patterson et al. 1995 and other references below). 
Those variations usually involve the appearance and growth of secondary humps 
which may -- occasionally -- replace the main maximum producing thereby 
a spurious phase shift in the (O-C) diagram; this is what happens in the case 
of the so-called late superhumps (cf. Udalski 1990, Patterson et al. 1995,1998).  

The aim of the present paper is to study the evolution of superhumps during 
the superoutbursts by decomposing their light curves into their Fourier 
components. This was done earlier by several authors, usually in order to prewhiten 
the observed curves (e.g. Olech et al. 2004a, 2004b, Otulakowska-Hypka 2013), 
but without paying attention to the results. 
And the results turn out to be important and revealing.

%Section 2
\section {The Data and their Analysis } 

We begin be describing the superhump light curves used in our analysis. 
They are listed in Table 1. 

%Table 1
\begin{table}[h!]
{\parskip=0truept
\baselineskip=0pt {
\medskip
\centerline{Table 1}
\medskip
\centerline{ The Superhump Light Curves }
\medskip
$$\offinterlineskip \tabskip=0pt
\vbox {\halign {\strut
\vrule width 0.5truemm #&	%1
\enskip\hfil#\enskip&	        %2
\vrule#&			%3
\enskip\hfil#\enskip&           %4
\vrule#&			%5
\enskip#\enskip\hfil&           %6
\vrule width 0.5 truemm # \cr	%7
\noalign {\hrule height 0.5truemm}
&&&&&&\cr
& Star \hfil && Superoutburst \hfil&& Source of light curves \hfil&\cr
&&&&&&\cr
\noalign {\hrule height 0.5truemm}
&&&&&&\cr
&   TT Boo &&      June 2004 && Olech et al. (2004b), Fig.3 &\cr
&   GZ Cnc &&     March 2010 && Kato et al (2010), Fig.9 &\cr
&   PU CMa &&  November 2009 && Kato et al (2010), Fig.4 &\cr
&   GX Cas &&   October 2010 && Kato et al (2012), Fig.7 &\cr
&   HT Cas &&  November 2010 && AAVSO International Database  &\cr
&   IX Dra && September 2003 && Olech et al (2004a), Fig.7 &\cr
&   IX Dra &&   October 2010 && Otulakowska-Hypka et al. (2013), Fig.10 &\cr
&   XZ Eri &&   January 2003 && Uemura et al. (2004), Fig.4 &\cr
& V344 Lyr &&    August 2009 && Wood et al. (2011), Fig.18 &\cr
&   ER UMa &&   January 1995 && Kato et al. (2003), Fig.4 &\cr
&   KS UMa &&  February 2003 && Olech et al. (2003), Fig.6 &\cr
&&&&&&\cr
\noalign {\hrule height 0.5truemm}
}}$$
}}
\end{table}

{\it TT Boo}. The superhump period variations were quite complex 
(see Fig.5 in  Olech et al. 2004b) and the average daily superhump light 
curves (their Fig.3) show secondary humps appearing after JD 2453169.  

{\it GZ Cnc}. The daily average superhump light curves clearly show 
a secondary hump which appeared already on the second night and grew in amplitude 
which on JD 2455275 became higher than that of the original maximum.  
The phases from Fig.9 of Kato et al. (2010) are corrected here by 
$\Delta\phi=0.25$ thereby making $\phi=0$ correspond to maximum. 

{\it PU UMa}. The light curves show a secondary hump starting with JD 2455162. 

{\it GX Cas}. The average daily superhump light curves are corrected for 
variable period. They show secondary hump starting with JD 2455502. 

{\it HT Cas}. B{\c a}kowska et al. (2014ab) and Kato et al. (2012) presented 
results describing the November 2010 superoutburst of this star but without any  
information on the evolution of the shapes of superoutburst light curves.  
The data used here were taken from the AAVSO International Database 
and analyzed in 1-day intervals. They clearly show the development of 
a secondary hump. 

{\it IX Dra (2003)}. The secondary hump was present already on the second night 
(JD 2452904). Three days later its amplitude was equal to, and on the last two 
nights (JD 2452908 and 909) was higher than that of the primary maximum. 
The modulation of the light curve was so strong that the Authors (Olech et al. 
2004a) used {\it minima} to determine the period and phases; consequently 
$\phi=0$ corresponds to minimum. 

{\it IX Dra (2010)}. The secondary hump appeared on the third night 
(JD 2455480). Around JD 2455802 the phase of the primary maximum showed a jump 
by $\Delta\phi\approx 0.15$ (see Fig.13 in Otulakowska-Hypka et al. 2013). 
The phases are corrected here by $\Delta\phi \approx$0.05/day in order to make 
$\phi_{max}\approx 0$ during the first four nights. 

{\it XZ Eri}. The secondary hump was present starting with JD 2452670. 

{\it V344 Lyr}. The Authors (Wood et al. 2011) determined the average superhump 
light curves in 5-day intervals with $\phi=0$ corresponding to minimum. 

{\it ER UMa}. The secondary humps were present starting with the first night. 
Around JD 2449747 the phase of the primary maximum showed a jump 
by $\Delta\phi\approx 0.5$ (see Fig.1 in Kato et al. 2003). 

{\it KS UMa}. Clear secondary humps did not appear until around JD2452690. 

Nearly all periodograms of superhumps which can be found in the literature 
are limited to the region of the fundamental mode and only seldom include 
the region of the first overtone. 
The only exception was the analysis of V344 Lyr by Wood et al. (2011,Fig.10), 
their discrete Fourier Transform extending up to the fifth overtone and  
showing significant power up to the third overtone. 
Consequently the light curves described above were decomposed into their 
fundamental mode and the first three overtones 

%Eq.1
\beq
m~=~<m>~-~\sum_{k=0}^3~2\pi~(k+1)~A_k~\cos (\phi-\phi_{max}^k) ,
\eeq 

\noindent 
where $k=0$ corresponds to the fundamental mode, while $k=1,2,3$ -- to the 
overtones. 
At this point the presence of higher overtones, with much smaller but 
non-negligible amplitudes, cannot be excluded. Therefore whenever a reference 
is made in the text to "three overtones" or "four periods" the reader will be 
expected to add "or more?".  

The analysis was limited to the superhump light curves observed during 
the main part -- plateau -- of a given superoutburst. 
Results are presented and discussed in the next two sections.

%Section 3
\section { The Amplitudes }

The behavior of amplitudes during superoutbursts is shown in Fig.1. 
Three sets of parameters are determined for each star: 
the average amplitude $<A_k>$, the maximum amplitude $A_{k,max}$ and the rate 
of decline $dA_k/dt$. 

%***Fig.1
\begin{figure}[htb]
\epsfysize=17.0cm 
\hspace{-1.5cm}
\epsfbox{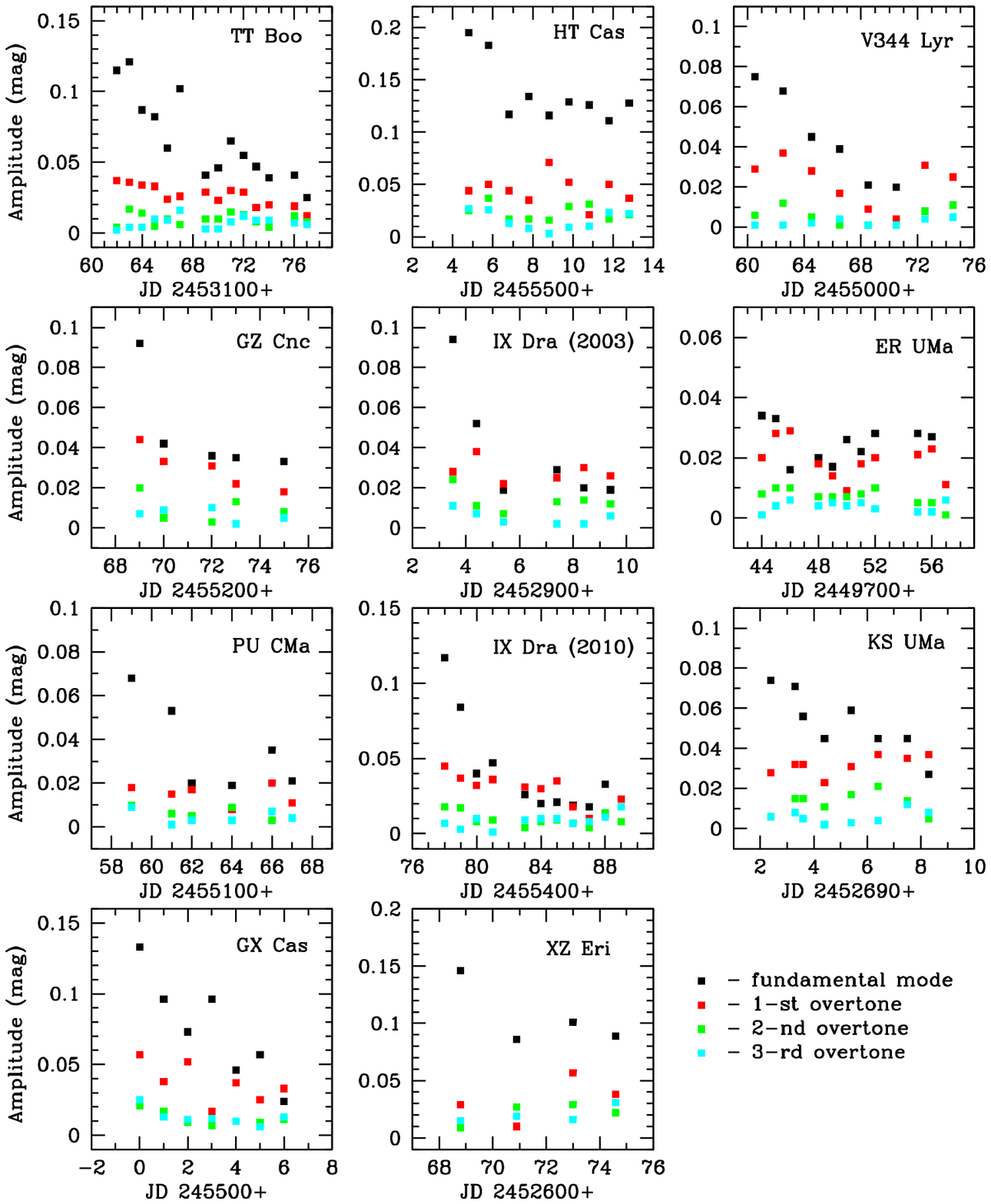} 
\vskip -10truemm
\FigCap { Evolution of amplitudes -- of the fundamental mode and the 
first three overtones --  during superoutburst. 
Errors (not shown) are comparable or only slightly larger than the size of the 
symbols. }
\end{figure}

%***Fig.2
\begin{figure}[htb]
\epsfysize=17.0cm 
\hspace{-1.5cm}
\epsfbox{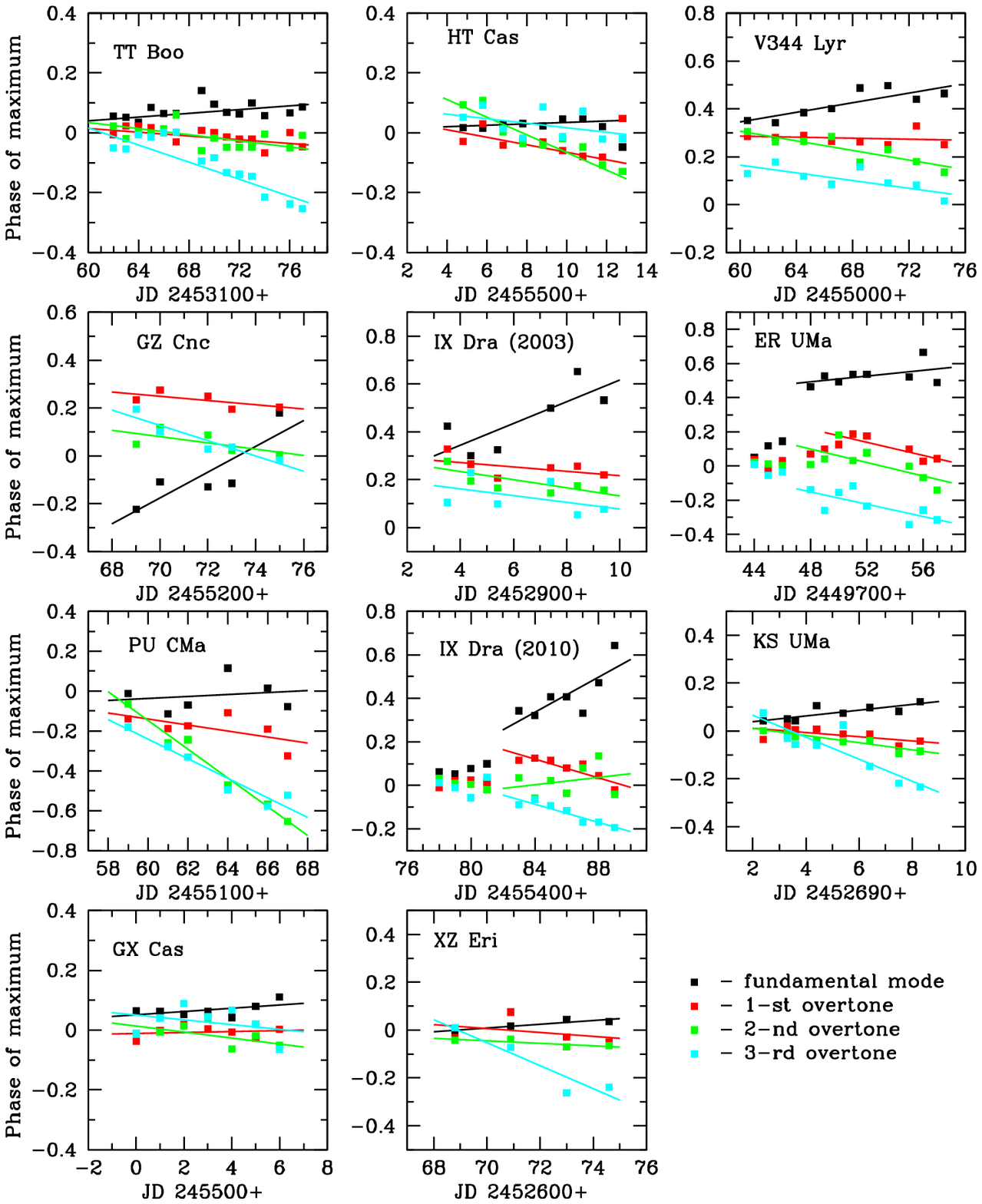} 
\vskip -10truemm
\FigCap { Evolution of the phases of maxima -- of the fundamental mode and the 
first three overtones -- during superoutburst. 
Errors (not shown) are comparable or only slightly larger than the size of the 
symbols. For special comments on IX Dra (2010) and ER UMa see Section 4.4. }
\end{figure}

The amplitudes of the overtones are much smaller than those of the fundamental 
mode which is illustrated by the following sets of values: 
$<A_k>=0.062$, 0.029, 0.012, 0.008 and $A_{k,max}=0.086$, 0.034, 0.014, 0.008 
for, respectively, $k=1,2,3,4$. It appears that the amplitudes of overtones 
higher than k=3 would be negligible. 

The amplitudes of the fundamental mode systematically decrease with time 
at the average rate $dA_0/dt=-0.0066\pm0.0013$ mag/day. 
Those of the overtones either decrease sligtly, remain practically constant, 
or even slightly increase (e.g. the 1-st overtone in KS UMa), the everage rates being: 
$dA_1/dt=-0.0012\pm0.0007$, $dA_2/dt=-0.0004\pm0.0003$ and 
$dA_3/dt=0.0000\pm0.0003$ mag/day. 

The obvious consequence is that at the begining of the superoutburst 
the superhump light curves are dominated by the fundamental mode. 
As the amplitude of the fundamental mode decreases and becomes  
comparable to those of the overtones, their contributions to the superhump 
light curve become significant.

%Section 4
\section {The Phases of Maximum } 

\subsection { Results }

The behavior of phases of maxima during superoutburst is shown in Fig.2. 
In all cases the phase of maximum of the fundamental mode {\it increases} 
with time while in practically all cases those of maxima of the overtones 
{\it decrease}. 

The values of $d\phi_{max}^k/dt$ were determined for all individual cases  
and the lines with slopes corresponding to those values are shown in Fig.2. 
Their average values are: $<d\phi_{max}^0/dt>=+0.018\pm0.006$, 
$<d\phi_{max}^1/dt>=-0.009\pm0.002$, $<d\phi_{max}^2/dt>=-0.017\pm0.006$, 
and $<d\phi_{max}^3/dt>=-0.024\pm0.005$. 
The differences between those values for $k=1,2,3$ are significant 
only at the $1\sigma$ level but not at the $3\sigma$ level. 
This could suggest that they are practically identical. 
Comparison of individual values of $d\phi_{max}^k/dt$ (Fig.3) shows, however 
that this is not the case. 

%***Fig.3
\begin{figure}[htb]
\epsfysize=10.0cm 
\hspace{1.5cm}
\epsfbox{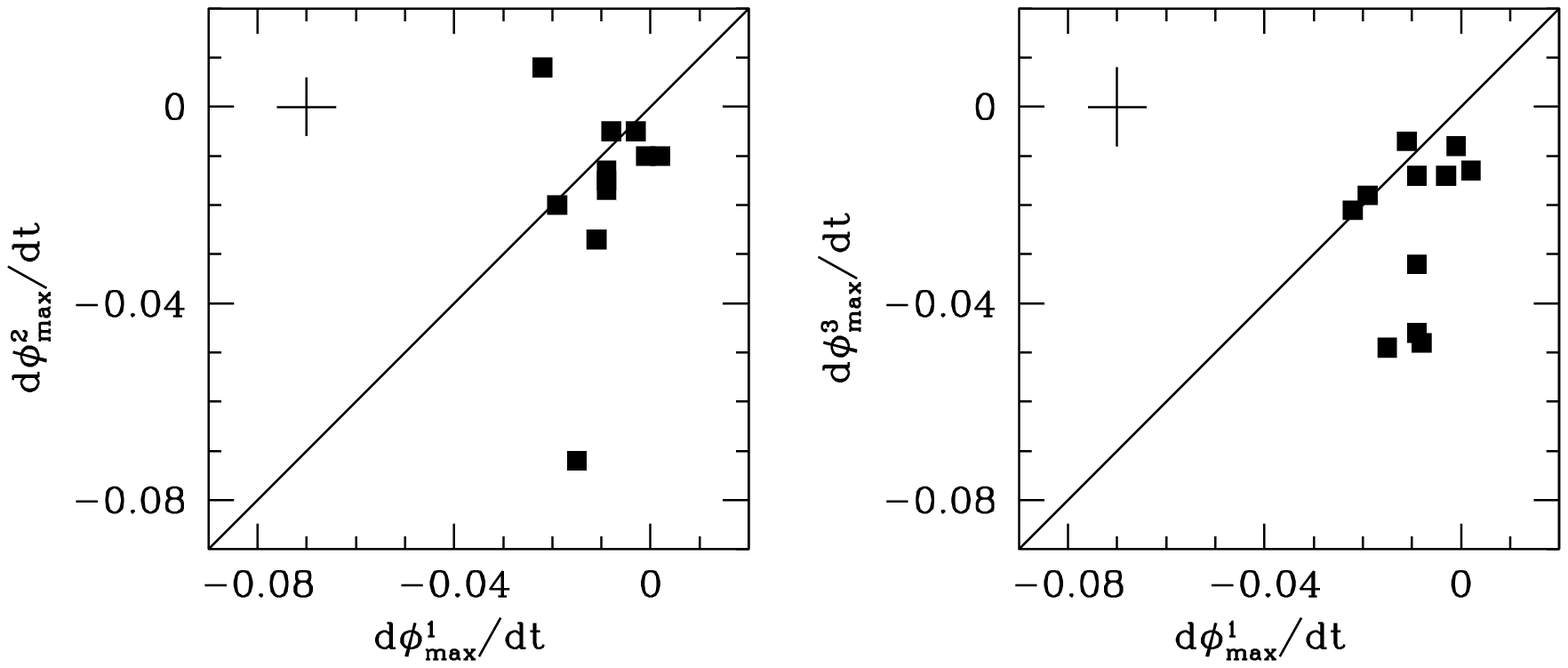} 
\vskip -50truemm
\FigCap { Comparison of individual values of $d\phi_{max}^k/dt$ for $k=1,2,3$. 
Typical errors are shown in the upper left corners. }
\end{figure}

\subsection {Phase Shifts versus Four Periods } 

There are two possible interpretations of the "phase diagrams" shown in Fig.2. 
The first is that we are dealing with time-dependent, genuine 
phase shifts (see Section 5). 

The second interpretation, based on an obvious analogy between the 
"phase diagrams" and the traditional (O-C) diagrams, is that instead of 
just one superhump period $P_{sh}$ there are four periods $P_k$. 
They resemble -- but are significantly different from -- the fundamental mode 
and the first three overtones of $P_{sh}$:   

%Eq.2
\beq
P_k~=~{1\over {k+1}}~\Bigl[~P_{sh}~+~P_{sh}^2~{{d\phi_{max}^k}\over{dt}}~\Bigr] ,
\eeq

\noindent
where $P_{sh}$ is the period used for calculating phases shown in Fig.2. 

One can also consider the corresponding periods of the fundamental mode 

%Eq.3
\beq
P_{k,fm}~=~(k+1)~P_k .
\eeq

The obvious question is: why those periods have not been detected earlier? 
The answer is simple. 
The peaks in the periodograms are generally too broad to distinguish between 
$f_k$ and $(k+1)\times f_{sh}$, which differ only very slightly. 
It may be also worth to add that the periodograms, which can be found in the 
literature, nearly always cover only the region around $f_{sh}$ and do not 
extend to higher frequencies.

\subsection { The Case of GZ Cnc and IX Dra (2003) } 

Those two examples illustrate how the presence and evolution of the secondary 
hump can produce spurious phase jumps in the (O-C) diagrams. 

As already mentioned in Section 2 the modulation of the light curves in those two 
cases was very strong. The amplitude of the secondary hump grew considerably and 
eventually exceeded that of the primary maximum. 
In GZ Cnc this happened on JD 2455275 while in IX Dra -- on JD 2452908  
(one day earlier the two amplitudes were already equal). 

%***Fig.4
\begin{figure}[htb]
\epsfysize=10.0cm 
\hspace{1.5cm}
\epsfbox{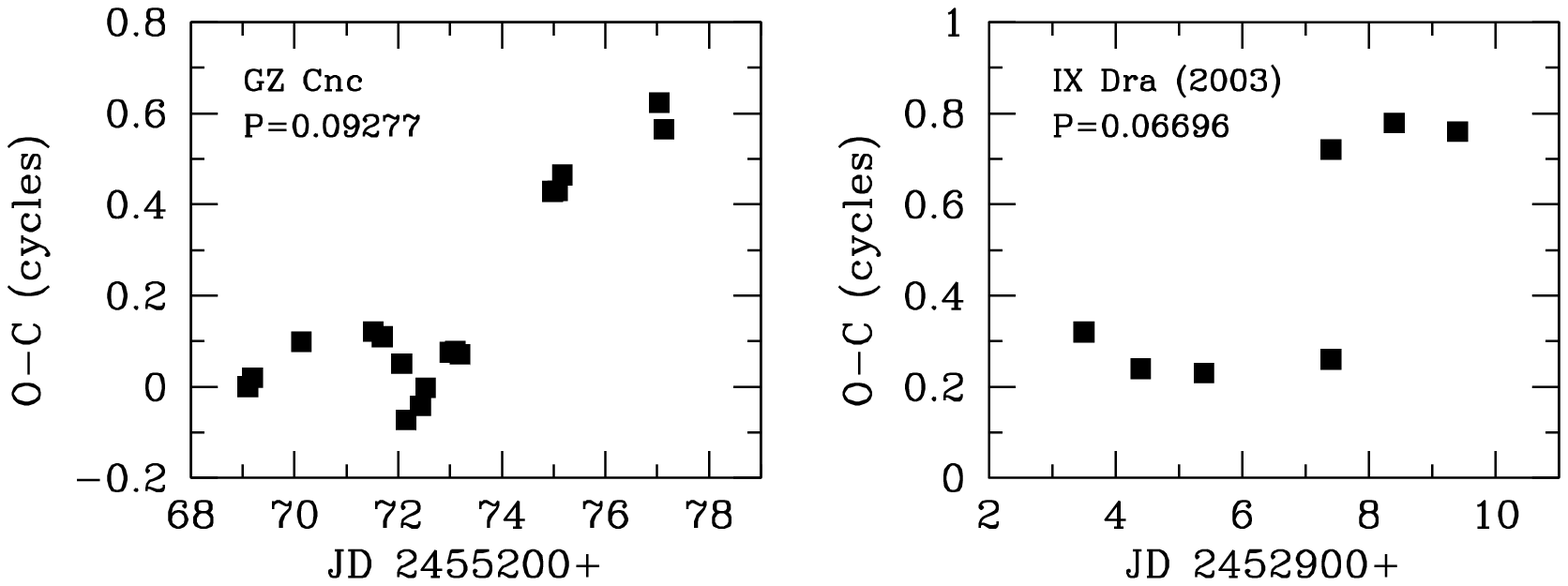} 
\vskip -60truemm
\FigCap { The (O-C) diagrams of GZ Cnc and IX Dra (2003) based on maxima of superhumps. }
\end{figure}

Fig.4 presents the (O-C) diagrams based on moments of maxima taken from 
-- respectively -- Table 9 in Kato et al. (2010) and Fig.7 in Olech et al. (2004a). 
As could be expected they show spurious phase jumps by $\Delta \phi \sim 0.4P$ 
and $\sim 0.5P$.

\subsection { The Case of IX Dra (2010) and ER UMa } 

Those two cases differ from the others and require additional comments. 
The (O-C) diagram of IX Dra (Otulakowska-Hypka et al. 2013, Fig.13) shows 
around JD 2455483 a phase jump by about $0.15P$. 
The phase diagram (Fig.2) shows that this was largely due to a phase jump in 
$\phi_{max}^0$ (there was also a smaller phase jump in $\phi_{max}^1$). 
After that event all four $\phi_{max}^k$ followed the "standard" pattern.
 
The case of ER UMa was similar: The (O-C) diagram (Kato 2003, Fig.1) 
shows around JD 2449747-48 a phase jump by about $0.5-0.6P$. 
The phase diagram (Fig.2) shows a similar phase jump in $\phi_{max}^0$ 
followed by the "standard" behavior of all four $\phi_{max}^k$. 

The nature of those genuine phase jumps of the fundamental mode component 
requires explanation.

%Section 5
\section { Discussion }  

Results presented in this paper create problems which complicate 
our understanding of superhumps. 

The first of them is related to the periods of superhumps. 
According to their commonly accepted interpretation they are related to 
the period of apsidal motion of the disk:  

%Eq.5 
\beq
{1\over{P_{sh}}}~=~{1\over{P_{orb}}}~-~{1\over{P_{aps}}} .
\eeq

Furthermore, it is commonly accepted that the period of disk's apsidal motion 
is a function of its radius: $P_{aps}=f(r_d)$ (cf. Montgomery 2001, Murray 2000 
and references therein). 

Replacing $P_{sh}$ in Eq.(5) by four {\it different} periods $P_{k,fm}$ 
would imply four different apsidal motion periods $P_{k,aps}$ and four 
different disk radii $r{k,d}$! Obviously then the "four-period" interpretation 
must be abandoned in favor of "genuine phase shifts".  

The commonly accepted tidal-resonance model (Whitehurst 1988, Hirose and Osaki 
1990; see also Smith et al. 2007 and references therein), explains superhumps 
as being due to tidal dissipation effects in the outer parts of accretion disk 
undergoing apsidal motion.  
As discussed earlier (Smak 2010) this model fails to explain many important facts.  
Whether and how it could explain the time-dependent phase shifts is a question 
addressed to its adherents.  

The irradiation modulated mass transfer model proposed on purely 
observational evidence by the present author (Smak 2009,2013) explains superhumps 
as being due to periodically modulated dissipation of the kinetic energy 
of the stream. Its essential ingredient is the non-axisymmetric structure of 
the outer parts of the disk involving the azimuthal dependence of their vertical 
thickness: $z/r=f(\alpha)$. 
It is quite possible that the behavior of phases of maxima $\phi_{max}^k$, 
seen in the phase diagrams, simply reflects the evolution of the disk outer 
structure and the shape of $f(\alpha)$ but -- for the time being -- this is 
only a speculation.

\Acknow{ It is the author's pleasant duty to acknowledge with thanks the 
use of the observations of HT Cas from the AAVSO International Database. 
The author is also grateful to Professor Aleksander Schwarzenberg-Czerny 
for a helpful discussion,
}

\begin {references} 

\refitem {B{\c a}kowska, K., Olech, A., Rutkowski, A., Koff, R., 
          de Miguel, E., Otulakowska-Hypka, M. } {2014a} 
         {\it Contrib. Astron. Obs. Skalnat{\' e} Pleso} {43} {325}

\refitem {B{\c a}kowska, K., Olech, A. } {2014b} {\Acta} {64} {247} 

\refitem {Hirose, M., Osaki, Y.} {1990} {\PASJ} {42} {135} 

\refitem {Kato, T., Nogami, D., Masuda, S. } {2003} {\PASJ} {55} {L7} 

\refitem {Kato, T. et. al.} {2009} {\PASJ} {61} {S395}

\refitem {Kato, T. et al. } {2010} {\PASJ} {62} {1525} 

\refitem {Kato, T. et al. } {2012} {\PASJ} {64} {21} 

\refitem {Montgomery, M.M.} {2001} {\MNRAS} {325} {761} 

\refitem {Murray, J.R.} {2000} {\MNRAS} {314} {L1} 

\refitem {Olech, A., Schwarzenberg-Czerny, A., K{\c e}dzierski, P.,  
          Z{\l}oczewski, K.,Mularczyk, K., Wi{\'s}niewski, M.} {2003} 
          {\Acta} {53} {175} 

\refitem {Olech, A., Z{\l}oczewski, K., Mularczyk, K.,K{\c e}dzierski, P., 
          Wi{\'s}niewski, M., Stachowski, G.} {2004a} {\Acta} {54} {57} 

\refitem {Olech, A., Cook, L.M., Z{\l}oczewski, K., Mularczyk, K., 
          K{\c e}dzierski, P., Udalski, A., Wi{\'s}niewski, M.} 
         {2004b} {\Acta} {54} {233} 

\refitem {Otulakowska-Hypka, M., Olech, A., de Miguel, E., Rutkowski, A., 
          Koff, R., B{\c a}kowska, K.} {2013} {\MNRAS} {429} {868} 

\refitem {Patterson, J., Jablonski, F., Koen, C., O'Donoghue, D., Skillman, D.R.} 
         {1995} {\PASP} {107} {1183} 

\refitem {Patterson, J. et al.} {1998} {\PASP} {110} {1290} 

\refitem {Patterson, J.} {1999} { {\it Disk Instabilities in Close Binary 
          Systems}, {\rm eds. S.Mineshige and J.C.Wheeler (Tokyo: Universal
          Academy Press)} } {~} {61}

\refitem {Smak, J.} {2009} {\Acta} {59} {121}

\refitem {Smak, J.} {2010} {\Acta} {60} {357}

\refitem {Smak, J.} {2013} {\Acta} {63} {369}

\refitem {Smith, A.J., Haswell, C.A., Murray, J.R., Truss, M.R., 
          Foulkes, S.B.} {2007} {\MNRAS} {378} {785}

\refitem {Still, M., Howell, S.B., Wood, M.A., Cannizzo, J.K., Smale, A.P.}
          {2010} {\ApJ} {717} {L113}

\refitem {Udalski, A.} {1990} {\AJ} {100} {226}

\refitem {Uemura, M. et al.} {2004} {\PASJ} {56} {141} 

\refitem {Vogt, N.} {1974} {\AA} {36} {369}

\refitem {Warner, B.} {1975} {\MNRAS} {170} {219} 

\refitem {Whitehurst, R.} {1988} {\MNRAS} {232} {35} 

\refitem {Wood, M.A., Still, M.D., Howell, S.B., Cannizzo, J.K., Smale, A.P.} 
         {2011} {\ApJ} {741} {105} 

\end {references}

\end{document}